\documentclass[conference]{IEEEtran}
\IEEEoverridecommandlockouts
\usepackage[dvipdfmx]{graphicx}
 \usepackage{lipsum,graphicx,subcaption}
\usepackage{float}
\usepackage{cite}
\usepackage{amsmath,amssymb,amsfonts}
\usepackage{algorithmic}
\usepackage{textcomp}
\usepackage{algorithmic,algorithm}
\usepackage{url}
\usepackage{color}
\usepackage[11pt]{moresize}
\def\BibTeX{{\rm B\kern-.05em{\sc i\kern-.025em b}\kern-.08em
    T\kern-.1667em\lower.7ex\hbox{E}\kern-.125emX}}

\DeclareMathOperator*{\argmax}{arg\,max}

\begin{document}

\title{Joint Power Allocation in Interference-Limited Networks via Distributed Coordinated Learning}

\author{
    \IEEEauthorblockN{Roohollah Amiri\IEEEauthorrefmark{1}, Hani Mehrpouyan\IEEEauthorrefmark{1}, David Matolak\IEEEauthorrefmark{2}, Maged Elkashlan\IEEEauthorrefmark{3}}
    \IEEEauthorblockA{\IEEEauthorrefmark{1}\textit{\small{School of Electrical and Computer Engineering, Boise State University,
    \{roohollahamiri,hanimehrpouyan\}@boisestate.edu}}}
    \IEEEauthorblockA{\IEEEauthorrefmark{2}\textit{\small{Department of Electrical Engineering, University of South Carolina, matolak@cec.sc.edu}}}
    \IEEEauthorblockA{\IEEEauthorrefmark{3}\textit{\small{School of Electronic Engineering and Computer Science, Queen Mary University of London, maged.elkashlan@qmul.ac.uk}}}
}

\maketitle

\begin{abstract}
Dense deployment of small base stations (SBSs) is one of the main methods to meet the 5G data rate requirements. However, high density of independent SBSs will increase the interference within the network. To circumvent this interference, there is a need to develop self-organizing methods to manage the resources of the network. In this paper, we present a distributed power allocation algorithm based on multi-agent Q-\textit{learning} in an interference-limited network. The proposed method leverages coordination through simple message passing between SBSs to achieve an optimal joint power allocation. Simulation results show the optimality of the proposed method for a two-user case.%The complexity of the proposed method is fixed, regardless of the fact that the search space of the system grows exponentially with the number of base stations in the network.
%In future cellular network, i.e.\ 5G, residential areas will be supported by user-mounted small base stations (SBSs). These SBSs work independently and make a heterogeneous network. Due to unplanned structure of the SBSs, co-channel interference is unavoidable, especially in high density residential areas. In these network, a self-organized, distributed resource allocation algorithm will be a necessity. 
\end{abstract}

%\begin{IEEEkeywords}
%Self-organization, dense network, Q-\textit{learning}, fast clustering.
%\end{IEEEkeywords}

\section{Introduction}
Ultra-densification through the use of smaller base stations is a promising technology in the next generation of cellular networks (5G)~\cite{art_dense_3}. The small base stations (SBSs) might be mounted by users in a plug-and-play fashion, and their backhaul may be supported by broadband connections. The user-mounted feature, introduces unplanned deployment of SBSs, which may result in unavoidable co-channel interference.

The problem of power allocation in an interference-limited network has been investigated widely in the literature. In~\cite{art_PA_0} and~\cite{art_PA_1}, the optimal power allocation for a two-user interference channel is derived for sum and individual power constraints, respectively. In~\cite{art_cvx_0} a more general solution is proposed for multi-transmitter systems with individual power constraints. The solution depends on the signal-to-interference-plus-noise ratio (SINR) value. In high SINR regime, the optimal solution is derived through transforming the problem into a geometric programming (GP) problem, while in the low SINR regime, a heuristic solution based on solving multiple GPs is used. It is important to note that all of these prior approaches are based on interior point methods. Hence, they require a centralized network management approach which may be impossible in dense networks. In~\cite{art_cvx_0}, a distributed method based on decomposing the optimization problem into local problems is proposed. The solution is based on message-passing and applies to high SINR case with full channel state information (CSI). Nonetheless, in a dense plug-and-play network, with a changing architecture, the assumptions of high SINR and the availability of full CSI at all nodes may not hold.%} {\color{red} WHY? AND WHICH METHODS? I WOULD CITE THEM AGAIN AND EXPLAIN WHY.}

In an ultra-dense network, in which the architecture of the network changes sporadically, a self-organizing method is a viable solution to manage the network resources. To this end, cooperative multi-agent reinforcement learning (MARL) methods have been used in resource management of communication networks~\cite{art_QL_dense_0,art_QL_dense_1,Amiri2018AML,amiri_gsmm,art_self_fuzzy}. Radio measurements such as SINR, are part of the Big data in cellular network~\cite{art_bigdata_0}, and one of the main advantages of MARL solutions is to utilize the measured SINR values. Generally most of the classic optimization solutions are based on channel coefficients. Thus, the prior methods require full CSI to find the solution while the MARL methods only need access to existing radio measurements, i.e., the measured SINR values. However, the existing MARL approaches in communication network management do not address the optimality of their cooperation methods. This is an important research topic to address since finding the optimal joint power allocation is directly impacted by the nature of the cooperation approach.% that is used.

%{\color{red} WHY IS THIS A BEFEIT? WHAT I MEAN IS WHY IS THE MEASURED SINR BEING MENTIONED HERE ONE WOULD EXPECT TO BE THE CASE.}{\color{blue} 	BECAUSE ALMOST, ALL OF THE SOLUTIONS IN CLASSIC SOLUTIONS ARE BASED ON CHANNEL COEFFICIENTS, HENCE THEY NEED CSI TO FIND THE SOLUTIONS. BUT MARL METHODS JUST USE AN EXISTING MEASUREMENT (SINR). IF YOU REMEMBER, RADIO MEASUREMENTS ARE PART OF THE BIG DATA IN CELLULAR NETWORKS AND HERE WE ARE TRYING TO USE THIS BIG DATA.} 

In this paper, we find an optimal joint power allocation solution via coordination between deployed SBSs. To address the optimality of the MARL approach, we model the whole system as a Markov decision process (MDP) with the SBSs being represented as the agents of the MDP. Subsequently, the value function of the MDP is approximated by a linear combination of local value functions of the SBSs. As we mentioned before, in order to remove the need for access to CSI, and develop an adaptable algorithm that handles a changing network architecture, each SBS uses a model-free reinforcement learning approach, i.e., Q-\textit{learning}. Q-\textit{learning} is used to update the SBS's local value function. Subsequently, we leverage the ability of SBSs to communicate over the backhaul network to build a simple message passing structure to coordinate them, based on variable elimination~\cite{art_coord_1}. Finally, we propose a distributed algorithm which finds an optimal joint power allocation in an interference-limited network.

%{\color{red} I AM NOT SURE FULLY HERE WHAT IS THE NOVELTY? THE WAY THIS READS IT APPEARS AS THOUGH YOU SIMPLY TOOK THE APPROACH IN THE PAPER CITED HERE AND APPLIED IT HERE. THIS DOES NOT SOUND GOOD EVEN IF IT IS TRUE AND MAYBE YOU CAN BE BIT MORE CLEAR ON WHAT IT IS THAT YOU ARE DOING THAT IS NEW AND SHOWCASE THE NOVELTY BETTER.} {\color{blue} THE MENTIONED REFERENCE, ~\cite{art_coord_1} , IS A COMPLETELY MATHEMATICAL PAPER, AND I AM USING THIS METHOD TO SOLVE A COMMUNICATION PROBLEM. THE IMPORTANCE OF THIS PAPER AS I HAVE MENTIONED, IS TO FIND OPTIMAL SOLUTION VIA MARL SOLUTION. THE OTHER WORKS DO NOT CARE ABOUT OPTIMALITY OF THEIR ANSWERS WHILE HERE WE PROPOSE AN ALGORITHM TO THAT.} {\color{red} IS THIS SOMETHING NEW? WHY USE Q-LEARNING? YOU NEED TO BE A BIT MORE INFORMATIVE ON YOUR CHOICES BECAUSE THEY WAY THIS IS WRITTEN IT APPEARS AS THOUGH YOU USED Q-LEARNING FOR NO APPARANT REASON.} {\color{red} OK THIS WHOLE PARAGRAPH DOES NOT INTER LAY ANY NOVELTY TO ME.}

% The idea of introduce a power allocation method based on coordination method in~\cite{art_coord_factored} which will achieve optimal joint allocation.

%\subsection{Motivation}
%Despite the progress in integration of reinforcement learning in resource management of wireless network, the optimality of these researches have not been addressed. In this paper we will show the direct impact of cooperation method on optimality of the solution. 

%\subsection{Related Work}

%\subsection{Contribution}

%\subsection{Paper Organization}
The paper is organized as follows. In Section~\ref{sec_systemModel}, the system model is presented. Section~\ref{sec_problem} first introduces the optimization problem, then analyzes the convexity of the problem. Section~\ref{sec_coord_qlearning} presents the general framework of the proposed solution. Section~\ref{sec_solution} outlines the proposed solution while Section~\ref{sec_sim} presents simulation results. Finally, Section~\ref{sec_con} concludes the paper.

\section{Network Model}\label{sec_systemModel}

%\subsection{Network Model}\label{sec_systemModel}

This paper considers downlink transmission in a dense deployment of $N$ small base stations (SBSs). We assumed each SBS supports one user equipment (UE), and all SBSs share the same frequency resource block. % {\color{red} CAN YOU JUSTIFY YOUR CHOICES HERE? IT ALWAYS HELPS THE READER. FOR EXAMPLE WHY DO YOU USE 1? WHY DO YOU USE THE SAME FREQUENCY, AND WHY IS THAT JUSTIFIED?} 
This system can represent a single cluster of a large network, which uses different frequency in each cluster to avoid interference between clusters. It is also assumed the SBSs are interconnected via a backhaul network supported by, for example, a broadband connection. Here, we use the same model of interference as~\cite{art_PA_0}. Thus, the received signal at the $i$th UE, $r_i$ is given by
{
\begin{align}\label{eq_signal}
r_i = \sqrt{g_i P_i} d_i + \sum_{j \in D_i} \sqrt{g_i P_j \beta_{ji}} d_j + n_i,
\end{align}
}
where $g_i$ represents the channel gain between the $i$th SBS and the UE it is serving, $d_i$ is the transmitted signal from the $i$th SBS, $P_i$ is the transmitted power at the $i$th SBS, $D_i$ represents the set of interfering SBSs to the $i$th UE, $\beta_{ji}~\left(0\leq \beta_{ji} \leq 1\right)$ for $1 \leq i \leq N$ and $j \in D_i$ is the ratio of the unintended power of the $j$th SBS when measured at the $i$th UE, and $n_i$ is the zero mean additive white Gaussian noise (AWGN) at the $i$th UE with variance $\sigma^2$.

%{\color{red} I SEE A HUGE PROBLEM HERE. ONE IS GOING TO ASK YOU HOW ARE THESE BETA'S MODELED. THE PROBLEM IS INTERFERENCE MODELING IS A BIG ISSUE HERE. ALSO IS THIS MMWAVE OR SUB-6 GHZ? I THINK SOME OF THESE ISSUES SHOULD BE ANSWERED HERE. ONE SOLUTION WOULD BE TO CORROBORATE YOUR EQUATION HERE AND SINR EXPRESSION BY ANOTHER PAPER THAT CONSIDERED A SIMILAR SCENARIO AND JUST CITE IT HERE.} {\color{blue} I HAVE USED THE SYSTEM MODEL IN \cite{art_PA_0}, BTW WE DO NOT ESTIMATE THE BETA HERE, SO WE DO NOT NEED TO WORRY ABOUT MODELING IT. IT'S SIMPLY A COEFFICIENT HERE WHICH SHOWS THE AMOUNT OF LEAKAGE FROM ONE CHANNEL TO ANOTHER. ALSO THIS APPROACH FITS MMWAVE AND SUB-6 GHZ BOTH. TH SYSTEM MODEL SUPPORTS BOTH OF THEM. TO USE ANY OF THEM WE JUST NEED TO USE THE CORRECT LARGE PATH LOSS MODEL TO DERIVE POWERS.}

According to the signal representation in ~\eqref{eq_signal}, the SINR at the $i$th UE, $\text{SINR}_i$, can be determined as
\begin{align}
\text{SINR}_i = \frac{g_i P_i}{\sum_{j \in D_i} g_i P_j \beta_{ji} + \sigma^2},
\end{align}
%with $\Gamma$ defined as the SINR gap which is selected according to the employed modulation, 
and the throughput at the $i$th UE normalized by the transmission bandwidth, $R_i$, is calculated as
\begin{align}\label{eq_throughput}
R_i = \log_2\left(1+\text{SINR}_i \right).
\end{align}
%It is worth noting that the proposed solution will use the measured SINR, and does not need to estimate the values of channel gains.
\subsection{Problem Analysis}\label{sec_problem}
Let us define $\mathbf{\underline{P}}=\left\lbrace P_1, P_2,...,P_N\right\rbrace$ as the set containing the transmitted power of the SBSs. The goal of the optimization is to find the optimal joint power allocation between SBSs, $\mathbf{\underline{P}^*}=\left\lbrace P_1^*, P_2^*,...,P_N^*\right\rbrace$, that maximizes the total throughput of the network. The optimization problem ($\textbf{OP}_1$) can be formulated as %{\color{red}WHAT ABOUT FAIRNESS? I JUST WANT TO KNOW ARE WE IGNORING IT? } {\color{blue} WE ARE NOT CONSIDERING FAIRNESS!}
\begin{subequations}\label{opt_1}
%\begin{eqnarray}
\begin{align}
& \underset{\underline{\mathbf{P}}}{\text{maximize}}
& & \sum_{i=1}^N R_{i} = \sum_{i=1}^N \log_2 \left(1+\text{SINR}_i\right),  \label{a}\\
& \text{subject to}
& & P_i \leq P_{i,max}, \; i = 1, \ldots, N.\label{b}
\end{align}
%\end{eqnarray}
\end{subequations}
Here, the objective function in \eqref{a} maximizes the sum throughput of the network. The constraint \eqref{b} refers to the individual power limitation of every SBS. %The term $\tilde{q}_k$ in \eqref{c} refers to the minimum required $\text{SINR}$ for the user $k$. The constraint in \eqref{c} ensures that the QoS is satisfied for all users. %Considering \eqref{eq_sinr_fue}, \eqref{eq_c2}, and \eqref{opt_1}, it can be concluded that the optimization in \eqref{opt_1} is a non-convex problem for dense HetNets. This follows from the SINR expression in \eqref{eq_sinr_fue} and the objective function \eqref{opt_1}. More specifically, the interference term due to the neighboring femtocells in the denominator of \eqref{eq_sinr_fue}, ensures that the optimization problem in \eqref{a} is not convex. This interference term may be ignored in low density networks but cannot be ignored in dense HetNets consisting of a large number of femtocells. 

%\subsection{Problem Analysis}\label{sec_probAnalysis}
%The optimization problem ($\textbf{OP}_1$) is a non-convex optimization problem. Here, first we will investigate the non-convexity of $\textbf{OP}_1$, and then we will examine the approximate solutions to the problem in two regimes : (1) high $\text{SINR}$, and low to medium $\text{SINR}$.
%\subsubsection{Non-Convexity of~$\textbf{OP}_1$}
The objective function in~\eqref{a} contains the interference term in the denominator of $\text{SINR}$ term. In a dense network the interference term cannot be ignored~\cite{art_slmz}. Due to the presence of the interference term, the objective function~\eqref{a} is a non-concave function~\cite{art_cvx_1}, which leads to non-convexity of the optimization problem.

\section{Distributed Coordinated Q-\textit{learning}}\label{sec_coord_qlearning}
In this section, the proposed optimal solution based on the Markov decision process (MDP) is presented. Then, the dimensionality issues of the optimal solution will be investigated. The dimensionality is important since it affects the tractability of the problem. Next, we use the coordination method introduced by~\cite{art_coord_1} to solve the problem in a distributed fashion. We show that the resulting method, provides a joint solution for the MDP via message passing between the agents of the network.

%The concept of coordination can be used to maximize the value function of a Markov decision process (MDP). Here, we first review the concept of value functions in MDP. Then we will cover how to solve the MDP through Q-\textit{learning}. After that, we will go through the idea of coordinating Q-\textit{learning} to achieve an optimal joint solution for the MDP.

\subsection{Optimal Solution via Q-\textit{learning}}
Consider a system with $N$ agents, where each agent $j$ selects its actions from its action set, $A_j$. Further, $\mathbf{X}=\left\lbrace X_1, X_2, ..., X_n\right\rbrace$ is the set of state variables which define the state of the system. Let us denote $\underline{x} \subset X$ to represent a single state of the system. In a fully cooperative game, we look for an optimal joint solution that is a Pareto optimal Nash equilibrium. One obvious solution to this problem is to model the whole system as a large MDP with its action set representing the joint action set of all the agents in the system. We consider $\mathbf{A}$ as the joint action set of all the agents, % $\left\lbrace A_1, A_2, ..., A_N\right\rbrace$ 
 and $\underline{a} \subset \mathbf{A}$ as a single joint action of this set.%, and $\mathbf{Q}$ as the total Q-function.

The MDP framework will be modeled as $\left(\mathbf{X}, \mathbf{A}, Pr, \mathbf{R} \right)$, where $\mathbf{X}$ denotes the finite set of states of the system, $\mathbf{A}$ is a finite set of joint actions, $Pr$ is the transition model which represents the probability of taking action $\underline{a}$ at state $\underline{x}$ and ending up in state $\underline{x}'$, $Pr\left(\underline{x},\underline{a},\underline{x}'\right)$, and $\mathbf{R}$ is the immediate reward received by taking action $\underline{a}$ at state $\underline{x}$, $\mathbf{R}\left( \underline{x},\underline{a}\right)$.

A policy, $\pi:\underline{x}\rightarrow \underline{a}$, for an MDP is defined as a strategy which shows at state $\underline{x}$, action $\pi\left(\underline{x}\right)$ will be taken. In order to evaluate a policy, a value function $V\left(\underline{x}\right)$, is defined which defines the value of policy at each state. In order to compute the value function for a given policy, we need to calculate the action-value function, also known as Q-function, defined as follows% INTERESTING SO MDPS WHERE USING Q-FUNCTIONS BEFORE? WAS THIS PRE DEFINED FOR THIS PROBLEM OR DID YOU HAVE TO DEFINE IT HERE? 
\begin{align}\label{eq_state_value_func}
\mathbf{Q}\left(\underline{x},\underline{a}\right) = \mathbf{R}\left(\underline{x},\underline{a}\right) + \gamma \sum_{\underline{x}'} Pr\left(\underline{x}'|\underline{x},\underline{a}\right) V\left(\underline{x}'\right),
\end{align}
in which $\gamma \in \left[ 0,1\right]$ is a discount factor. The optimal value at state $\underline{x}$ is the maximum value that can be reached by taking any action at this state. The optimal value function $V^*$, which gives the optimal policy $\pi^*$, satisfies the Bellman operation as follows~\cite{book_sutton}% DO WE NEED A REFERENCE HERE JUST SO PEOPLE KNOW WHERE TO GO TO READ MORE?
\begin{align}
V^*\left(\underline{x}\right) = \max_{\underline{a}} \mathbf{Q}^*\left(\underline{x},\underline{a}\right).
\end{align}

Q-\textit{learning} is a model-free reinforcement learning, which solves the Bellman equation through direct observations without knowledge of the transition model. In Q-\textit{learning}, the agent observers the state, $\underline{x}$, takes an action, $\underline{a}$, receives a reward, $\mathbf{R}$, and ends in a next state, $\underline{x}'$. Then, it will update its Q-function as follows
{
\begin{align}\label{eq_q_learning}
\mathbf{Q}\left(\underline{x},\underline{a}\right) = \mathbf{Q}\left(\underline{x},\underline{a}\right) +
\alpha [ \mathbf{R}\left(\underline{x},\underline{a}\right) + \gamma \max_{\underline{a}'} \mathbf{Q}\left(\underline{x}',\underline{a}'\right) - \mathbf{Q}\left(\underline{x},\underline{a}\right) ],
\end{align}
}where, $\alpha$ is the learning rate of the algorithm. If any action-state pair is repeatedly visited, the Q-function will converge to the optimal value~\cite{Watkins1992}.% In a multi-agent system, each agent $j$, will hold its local Q-function, $Q_j$, and the agents will choose their action by choosing the joint action that maximizes the joint Q-function.%, $\sum_j Q_j$. 
%In large MDPs, the action set $A$ can be a joint action of many agents. The size of a joint action set can grow exponentially with respect to number of agents, which makes the computation of $Q_V$ expensive. In order to solve this issue, according to~\cite{art_coord_1}, $Q_V$ can be approximated by the summation of local Q-functions of the agents as $\sum_j Q_j$. So by solving the local Q-functions of each agent, the general Q-function can be approximated. The Q-function for each agent can be computed using Q-\textit{learning}. 

One issue with this method is that the size of the joint action set is exponential with respect to the number of agents. If there are $N$ agents in the network, and each one has $|A|$ number of actions as the size of their action set, the size of the joint action set, $|\mathbf{A}|$, will be $|A|^N$. The exponential size of the joint action set makes the computation of the Q-function expensive and in most cases intractable.

\subsection{Factored MDP}
In most cases, for both representational and computational advantages, the state and action sets of an MDP can be factored into subsets based on the structure of the problem~\cite{art_factor_MDP}. In large MDPs, the global Q-function can be approximated by the linear combination of local Q-functions, i.e. {\small $\mathbf{Q}=\sum_j Q_j(\underline{a_j})$}~\cite{art_coord_1}. The $j$th local Q-function, $Q_j$, has the joint action set which is a subset of the global joint action set, $\mathbf{A}$. Here, we will define the joint action set of $Q_j$ by {\small $Scope\left[ Q_j\right] \subset\mathbf{A}$} for which $\underline{a_j}$ is a single joint action of this set.

\textit{In a communication network, each SBS plays the role of an agent in the multi-agent network. The action of SBS $j$, is the transmit power, $P_j$, that is used to transmit its signal to the intended user. From this point, an agent in a communication network, refers to the SBS. Generally, in wireless communication systems, each access point receives interference from specific local access points. Therefore, the approximation of global Q-function by linear combination of local Q-functions, applies to interference-limited communication networks.}

%WOW JUST PERFECTLY WRITTEN. SO NICE AND CLEAR. BTW LETS SUBMIT THIS WORK STRAIGHT TO A JOURNAL WITH SOME MINOR CHANGES AND MORE RESULTS. THIS IS WELL THOUGHT OUT! GOOD WORK
\subsection{Decomposition of Global Q-function}
The decomposition of the global Q-function, relies on the dependencies between the agents of the network. These dependencies can be represented by \textit{coordination graphs} (CGs)~\cite{art_coord_1}. Generally, there are two decomposition methods: agent-based and edge-based. The agent-based decomposition provides a suitable architecture for a distributed system with exact solution, while the edge-based decomposition is recommended for CGs with densely connected nodes~\cite{art_sparse_qlearning} and provides suboptimal solution. In this paper we will choose the agent-based decomposition since we are focused on achieving the optimal solution.% IS THERE ANOTHER REASON ASIDE FROM THIS. I MEAN I UNDERSTAND WHY YOU USE AGENT BASED METHODS INTUITIVELY BUT ANY MORE EXPLANATION WOULD BE HELPFUL. 

\begin{figure}[h!]
    \centering
    \begin{minipage}{0.45\columnwidth}
        \centering
        \includegraphics[width=0.9\columnwidth]{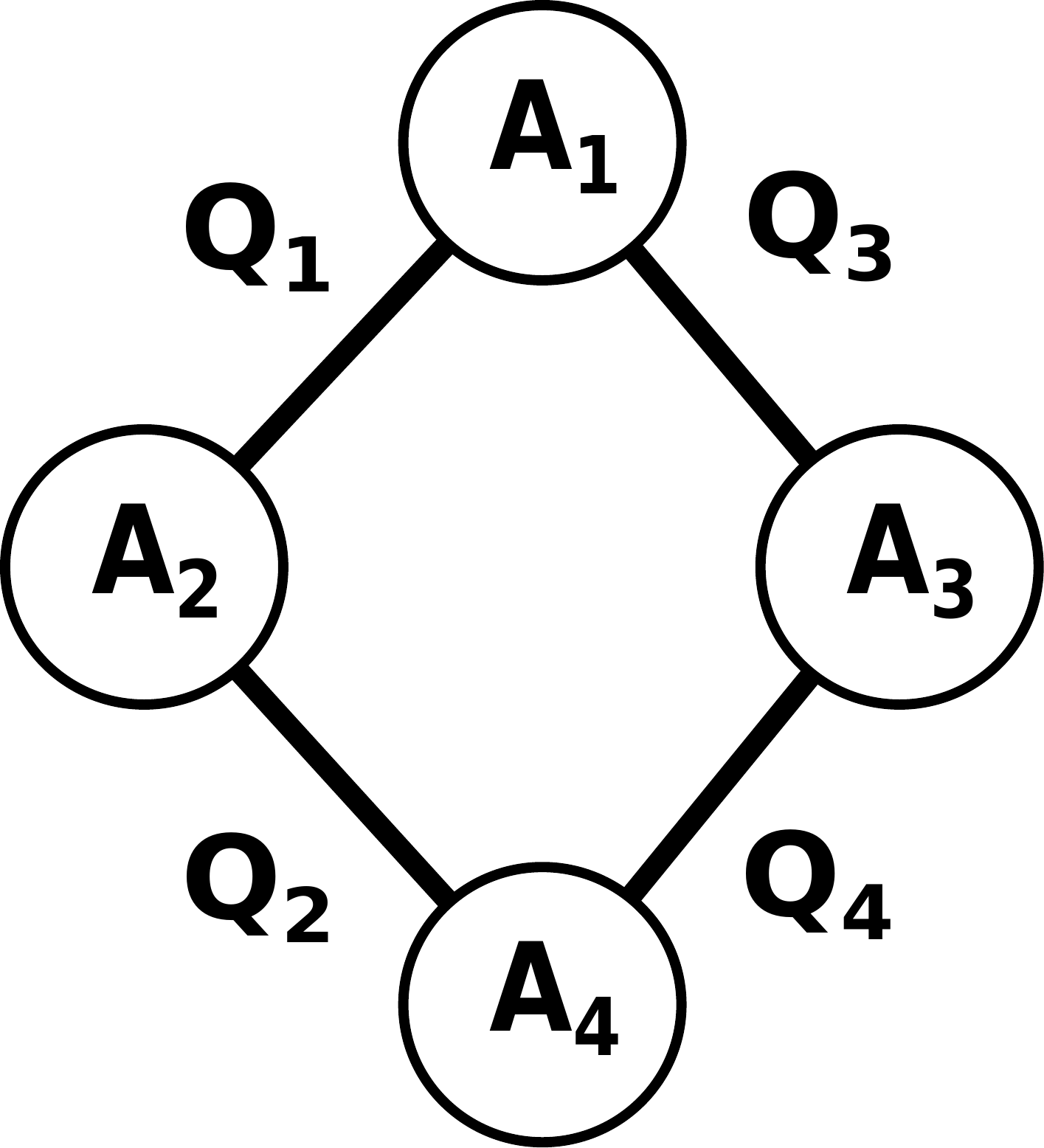}
		\caption{\small Coordination graph.}\label{fig_coord}
    \end{minipage}\hfill
    \begin{minipage}{0.45\columnwidth}
        \centering
        \includegraphics[width=0.9\columnwidth]{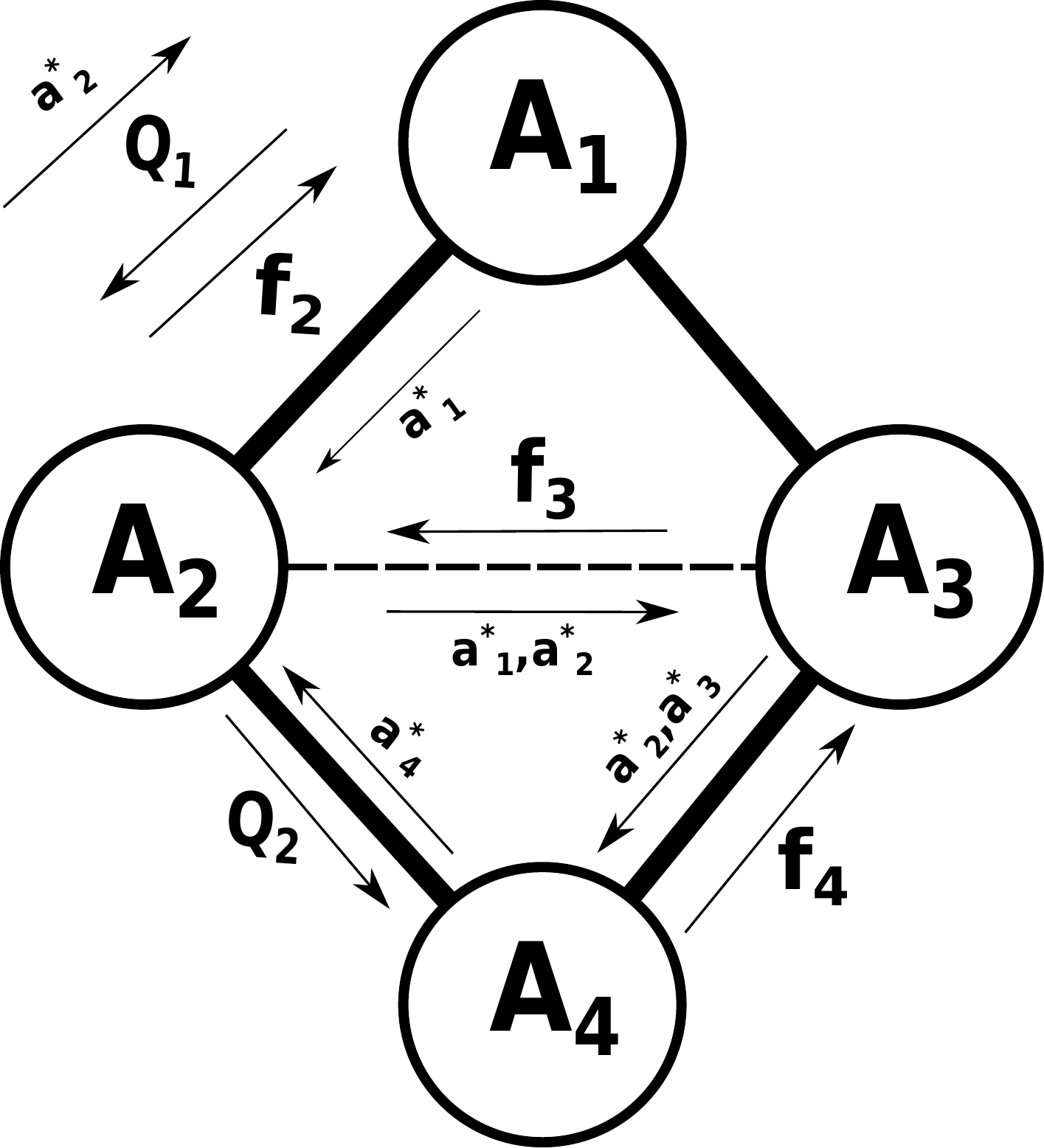} % first figure itself
        \caption{\small Message passing.}\label{fig_messages}
    \end{minipage}
\end{figure}

In a wireless network, the {\small $Scope\left[ Q_j\right]$} for agent $j$, is determined based on the interference model of the system, which is related to set $D$ in~\eqref{eq_signal}. For example, in Fig.~\ref{fig_coord}, four agents interfere with each other. Assume that agent $A_1$, receives interference from $A_2$ and $A_3$, and $A_4$ receives its interference from $A_2$ and $A_3$. Based on this model, the CG of the system is shown in Fig.~\ref{fig_coord}. Each edge between agents, shows a dependency between the two agents.% In this paper, based on the mmWave assumption, and the fact that mmWave band has a short range, we assume that each node in a CG is not densely connected to its local nodes, and each Q-function is at most depending on two agents. So the agent-based decomposition is chosen.

Here, we assume that all agents have the same state $\underline{x}$, hence, $Q\left(x,a\right)$ is written as $Q\left(a\right)$. According to the CG in Fig.~\ref{fig_coord}, the global Q-function, $\mathbf{Q\left(\underline{a}\right)}$, can be written as

{\small
\begin{align}
\mathbf{Q(\underline{a})} = Q_1(a_1,a_2) + Q_2(a_2,a_4) + Q_3(a_1,a_3) + Q_4(a_3,a_4).
\end{align}
}

%\vspace{-10pt}
\subsection{Coordinated Action Selection}\label{sec_coordination}

In multi-agent Q-\textit{learning}, according to~\eqref{eq_q_learning}, the agents will choose a joint action that maximizes the global Q-function. By using the agent-based decomposition, the joint action selection at state $\underline{x}$, $\max_{\underline{a}} \mathbf{Q}\left(\underline{a}\right)$, is written as

{\small
\begin{align}
\max_{a_1,a_2,a_3,a_4}Q_1(a_1,a_2) + Q_2(a_2,a_4) + Q_3(a_1,a_3) + Q_4(a_3,a_4).
\end{align}
}
This maximization problem, can be solved via variable elimination (VE) algorithm, which is basically similar to variable elimination in a Bayesian network~\cite{art_coord_factored}. Here, we will review this method for the network in Fig.~\ref{fig_coord}. The key idea is to maximize over one variable at a time, find conditional solutions, passing conditional functions to other agents, and sending back the results of local optimization to the related agents to recover their joint action choices.

We start from agent $A_4$. $a_4$ influences $Q_2$ and $Q_4$, so the maximization problem can be written as%  HOW DO YOU SHOW OPTIMALITY? I MEAN WHEN THINGS ARE NOT DONE JOINTLY USUALLY YOU DONT ACHIEVE OPTIMALITY.
%\vspace{-3pt}
{\small
%\begin{align}
%\max_{a_1,a_2,a_3}Q_1\left(a_1,a_2\right) + Q_3\left(a_1,a_3\right) + \left[ \max_{a_4} Q_2\left(a_2,a_4\right) + Q_4\left(a_3,a_4\right)\right].
%\end{align}
\begin{align}
\max_{a_1,a_2,a_3}Q_1(a_1,a_2) + Q_3(a_1,a_3) + [ \max_{a_4} Q_2(a_2,a_4) + Q_4(a_3,a_4)].
\end{align}
}
Agent $A_2$ communicates $Q_2$ to $A_4$, and $A_4$ solves its local maximization, which results in two functions: $f_4\left(a_2,a_3\right)$, and $b_4\left(a_2,a_3\right)$. These functions are defined as follows
{\small
\begin{align}
f_4\left(a_2,a_3\right) = \max_{a_4} Q_2\left(a_2,a_4\right) + Q_4\left(a_3,a_4\right),
\end{align}
\begin{align}
b_4\left(a_2,a_3\right) = \argmax_{a_4} Q_2\left(a_2,a_4\right) + Q_4\left(a_3,a_4\right).
\end{align}
}
At his stage, the $A_4$ has a conditional solution for $a_4$ based on $a_2$, and $a_3$, represented as the function $b_4$. Therefore, $A_4$ keeps $b_4$ and sends $f_4$ to its connecting agent, $A_3$. Then, $A_4$ is removed from the CG, and the maximization problem is translated to
%I SEE A HUGE PROBLEM HERE WE REALLY NEED SOMETHING HERE THAT SHOWS THIS IS OPTIMAL. I AM NOT SURE HOW WHEN YOU MAXIMIZED OVER THESE YOU CAN REMOVE THE EDGE AND JUST MOVE ON 
{\small
\begin{align}
\max_{a_1,a_2,a_3}Q_1\left(a_1,a_2\right) + Q_3\left(a_1,a_3\right) + f_4\left(a_2,a_3\right),
\end{align}
}$f_4$ brings a new edge in the coordination graph, an induced edge, which is shown with dashed line between $A_2$ and $A_3$ in Fig.~\ref{fig_messages}. %SHOULDN'T YOU REMOVE THE EDGES THAT YOU ARE DONE WITH FROM THIS FIGURE?
 The next agent to be removed is $A_3$. The maximization problem is rewritten as
{\small
\begin{align}
\max_{a_1,a_2}Q_1\left(a_1,a_2\right) + \left[\max_{a_3} Q_3\left(a_1,a_3\right) + f_4\left(a_2,a_3\right)\right].
\end{align}
}
With the same procedure, $A_3$ introduces $f_3\left(a_1,a_2\right)$, and $b_3\left(a_1,a_2\right)$. Accordingly, the problem reduces to
{\small
\begin{align}
\max_{a_1,a_2}Q_1\left(a_1,a_2\right) + f_3\left(a_1,a_2\right).
\end{align}
}
Next agent to choose its action is $A_2$, for which the problem results in
{\small
\begin{align}
f_1 = \max_{a_1} f_2\left(a_1\right),
\end{align}
}where, $f_2\left(a_1\right) = \max_{a_2} Q_1\left(a_1,a_2\right) + f_3\left(a_1,a_2\right)$. Finally, $A_1$ chooses its action based on maximizing the function $f_2\left(a_1\right)$. The results at this stage are $f_1$, and $a_1^*$. $f_1$ represents the maximum value of the global Q-function over $a_1, a_2, a_3$, and $a_4$, and $a_1^*$ is the optimal joint action for $A_1$. To recover the joint action choices, $A_1$ sends $a_1^*$ to $A_2$. Then $A_2$ chooses its action, $a_2 = b_2(a_1^*)$, and sends $a_1^*,a_2^*$ to $A_3$. $A_3$ and $A_4$ will choose their actions with the same procedure, $a_3^* = b_3(a_1^*,a_2^*)$, and $a_4^* = b_4(a_2^*,a_3^*)$.

In general, the elimination algorithm maintains a set of functions in each step. It starts with all local functions, $\left\lbrace Q_1, Q_2, ..., Q_N\right\rbrace$, and eliminates agents one by one.
%I UNDERSTAND A BIT MORE NOW BECAUSE YOU DO SEND IT BACK TO FIND THE OPTIMAL ONES. I THINK YOU MAY WANT TO PUT A NOTE IN THE EARLIER SECTION TO HELP THE READER SEE THIS BEFORE THEY REJECT THE PAPER AND NOT EVEN READ THIS SECTION. 

%In general, the elimination algorithm maintains a set of functions in each step, $\mathbf{F}$. It starts with all local functions, $\left\lbrace Q_1, Q_2, ..., Q_N\right\rbrace$. The algorithm steps can be summarized as follows
%\begin{enumerate}
%\item Choose an uneliminated agent, for example $A_l$.
%\item Choose all functions, $f_1, f_2, ..., f_l \in \mathbf{F}$ whose \textit{Scope} contains $A_l$.
%\item Define a new function, $f_l=\max_{a_l} \sum_j f_j$ and add it to $\mathbf{F}$. The \textit{Scope} of $f_l$ is $\cup_{j=1}^L~\text{Scope}\left[f_j\right]-\left\lbrace A_l \right\rbrace$.
%\end{enumerate}

\subsection{Local Update Rule}
After finding the joint action, each agent will update its local Q-function. The update rule in~\eqref{eq_q_learning} can be written as
%Since the global Q-function is approximated as the linear combination of local Q-functions, 
{\small
\begin{multline}\label{eq_update_rule1}
\sum_j Q_j\left(\underline{x},\underline{a_j}\right) = \sum_j Q_j\left(\underline{x},\underline{a_j}\right) + \\ \alpha \left[ \sum_j R_j\left(\underline{x},\underline{a_j}\right) + \gamma \max_{\underline{a}} \sum_j Q_j\left(\underline{x}',\underline{a}'\right) - \sum_j Q_j\left(\underline{x},\underline{a_j}\right) \right],
\end{multline}
}
where, the joint maximization is solved through VE according to the last section. %cannot be written as an independent summation of local functions, since it is a global joint optimization. Hence, the joint maximization should be solved through VE algorithm. 
By assuming $\underline{a}^*$ as the solution to the VE, and $\underline{a_j}^* \subset \underline{a}^*$ as the optimal joint action set for $Q_j$, the update rule for each local Q-function can be derived as

%I WAS VERY CONFUSED BY THIS WHOLE PARAGRAPH. I AM COMPLETELY LOST. PLEASE REWRITE IN A MORE COORDINATED WAY. 
{\small
\begin{align}\label{eq_update_rule2}
Q_j(\underline{x},\underline{a_j}) = Q_j(\underline{x},\underline{a_j}) + \alpha [ R_j(\underline{x},\underline{a_j}) + \gamma Q_j(\underline{x}',\underline{a_j}^*) - Q_j(\underline{x},\underline{a_j}) ].
\end{align}
}
The Fig.~\ref{fig_messages} illustrates all messages passed between the agents to solve VE and update local Q-functions.

%I SEE A HUGE PROBLEM HERE AGAIN. LOOK AT THE INTRODUCTION SECTION WHERE YOU SAY SECTION III REVIEWS THE NEEDED MATERIAL. HOWEVER THAT IS NOT TRUE. YOU ARE ACTUALLY OUTLINING YOUR PROPOSED SOLUTION. THE WAY THIS WAS WRITTEN, IT FEELS LIKE I AM UP TO PAGE 4 OF THE PAPER AND NOTHING NEW HAS BEEN PROPOSED JUST REVIEWING OTHERS WORK. THIS IS PARTIALLY TRUE BUT THE WAY YOU HAVE BROUGHT THESE WORKS TOGETHER IS UNIQUE AND YOU DID SO WHEN CONSIDERING THE CONTEXT OF WIRELESS NETWORKS. I CHANGED THE INTRODUCTION. SEE AND ACCEPT OR IGNORE ACCORDINGLY. 
\section{Power Allocation Using Coordinated Q-\textit{Learning} (Q-CoPA) }\label{sec_solution}
To integrate the idea of coordinated multi-agent learning into a communication network, we will model the SBS as an agent, and the whole network as a multi-agent MDP. The goal of the agents is to maximize total throughput of the network, as a cooperative game. 
\subsection{Q-CoPA Algorithm}
The proposed solution of this paper, \textit{Q-CoPA}, can be summarized as follows\\
\textit{The interference model of the network will be used to derive the coordination graph of the agents. The entire network is modeled as an MDP, and the global Q-function of the MDP is approximated by linear combination of local Q-functions of the agents. Each agent, based on the coordination graph, knows its \textit{Scope}. Local Q-functions are learned by the agents using cooperative Q-\textit{learning}. The cooperation method between the agents is to maximize the summation of local Q-functions by choosing an appropriate joint action. This action selection is implemented using variable elimination and message passing between the agents. The backhaul of the network is used as the infrastructure of message passing.}

The proposed method is represented in Algorithm 1.
\begin{algorithm}
\renewcommand\thealgorithm{}
\caption{1 The proposed Q-CoPA algorithm}\label{alg_1}
\begin{algorithmic}[1]
{\small
\STATE Initialize $\underline{x}$
\STATE Initialize All $Q_j(\underline{x},\underline{a_j})$ arbitrarily
\FORALL{episodes}
\STATE Choose $\underline{a}^*$ according to VE
\FORALL{agents}
\STATE Take action $\underline{a_j}$, observe $R_j$
\ENDFOR
\STATE Observe $\underline{x}'$
\STATE Calculate $\max_{\underline{a}'} \mathbf{Q}$ according to VE
\FORALL{agents}
\STATE Update local Q-function according to Eq.~\ref{eq_update_rule2}
\ENDFOR
\STATE $x_j \leftarrow x_j^{'}$
\ENDFOR
}
\end{algorithmic}
\addtocounter{algorithm}{-1}
\end{algorithm}

In the Algorithm 1, the loops at lines 5 and 10 are independent, and will be executed in parallel by the agents.
%where, $r_j$ is the immediate reward of agents $j$ at each episode, and $x_j^{'}$ is the next state observed by agent $j$ by taking action $a_j$.

\subsection{Q-\textit{learning} Parameters}
In the following the actions, and the reward of the Q-\textit{learning} algorithm implemented by each agent is defined.
\begin{itemize}
\item \textit{Actions} : Each SBS has a set of actions, which is defined as the transmit power levels. We define this set as $\left\lbrace p_1, p_2, ..., p_{N_{power}}\right\rbrace$. The number of power levels is defined as $N_{power}$.
\item \textit{Reward} : In each episode, SBS chooses a power level, and transmits its data to its intended user. The user measures the \text{SINR} of the signal, and will feedback it to the SBS. Then the reward of the SBS $j$ is calculated as $r_j = \log_2\left(1+\text{SINR}_j \right)$.
\end{itemize}

%I REALLY LIKE THE FINAL SOLUTION HERE. IT IS VERY ELEGANT. I AM CONCERNED ABOUT COMPLEXITY WHICH HAS NOT BEEN MENTIONED YET BUT HOPEFULLY IT COMES NEXT. MAYBE YOU CAN ALLUDE TO THIS SO PEOPLE DONT THINK YOU HAVE NOT THOUGHT ABOUT IT AT ALL. 

%I SEE ANOTHER BIG PROBLEM. YOU HARP ABOUT OPTIMALITY  SO MUCH B UT THERE WAS NO DISCUSSION AT ALL WHY THIS ALGORITHMS WILL GIVE US THE OPTIMAL SOLUTION. YOU NEED A REMARK HERE TO JUSTIFY AND EXPLAIN THIS. A SECOND REMARK ON COMPLEXITY. 
\section{Simulation Results} \label{sec_sim}
%In this section the proposed algorithm will be applied in a two-user case. The results of the solution will be compared to the optimal, greedy, and simultaneous solutions.

%\subsection{Simulation Setup}\label{sec_setup}
We consider two SBSs, each supporting one UE, with interfering channels. Each transmitter has omni-directional antenna and separate power source. The channel model is assumed to be time-invariant, i.e. slow fading. The channel gains are assumed to be $g_1=2.5$, and $g_2=1.5$. The $P_{1,max}=10$ dBm, $P_{2,max}=13$ dBm, and $\sigma^2=0$ dBm. Without loss of generality we assume that $\beta_{1,2}=\beta_{2,1}=\beta$ in Eq.~\ref{eq_signal}. % Quadrature amplitude modulation (QAM) is employed, therefore $\Gamma \equiv -\ln(5P_E)/1.5 $. 
The objective of the optimization is to find the power allocation to maximize the sum throughput of the network under individual power constraints.% The rest of the simulation parameters are provided in Table~\ref{table_params}.
%
%\begin{table}[h]
%\centering
%\caption{Simulation Parameters}
%\label{table_params}
%\begin{tabular}{|c|c|c|c|c|c|}
%\hline
%\textbf{Parameter} & \textbf{Value} & \textbf{Parameter} & \textbf{Value} & \textbf{Parameter} & \textbf{Value} \\ \hline
%$g_1$  & 2.5   &  $P_{1,max}$   & 10 dB   & $\sigma^2$ & 1  \\ \hline
%$g_2$  & 1.5   &  $P_{2,max}$   &  20 dB   & $P_E$      &  $10^{-4}$  \\ \hline
%\end{tabular}
%\end{table}

In executing the Q-CoPA algorithm, each Q-function is defined as a table, Q-table. The learning rate is $\alpha=0.5$, the discount factor as $\gamma=0.9$, $N_{\text{power}} = 100$, and the maximum number of episodes is set to $50$ times the size of a Q-table. The MDP of this problem is assumed to be stateless. The actions of agents are the transmit powers, $a_1=P_1$, and $a_2=P_2$, Q-functions are defined as: $Q_1(P_1,P_2)$ and $Q_2(P_1,P_2)$, and the global Q-function is defined as: {\small $\mathbf{Q}\left(P_1,P_2\right)=Q_1(P_1,P_2)+Q_2(P_1,P_2)$}.

According to~\cite{art_PA_1}, the optimal power allocation to maximize the sum-rate of the above network is derived as
{\small
\begin{equation}
  (P_1^*, P_2^*)=
  \begin{cases}
    (P_{1,max},0),~~~\text{if~} g_1 P_{1,max} \geq max\left(g_2 P_{2,max},1/\beta^2\right),\\
    (0,P_{2,max}),~~~\text{if~} g_2 P_{2,max} \geq max\left(g_1 P_{1,max},1/\beta^2\right),\\
    (P_{1,max},P_{2,max}),~~~\text{otherwise}.
  \end{cases}
\end{equation}
}

\begin{figure}[h]
\begin{centering}
\includegraphics[width=.90\columnwidth]{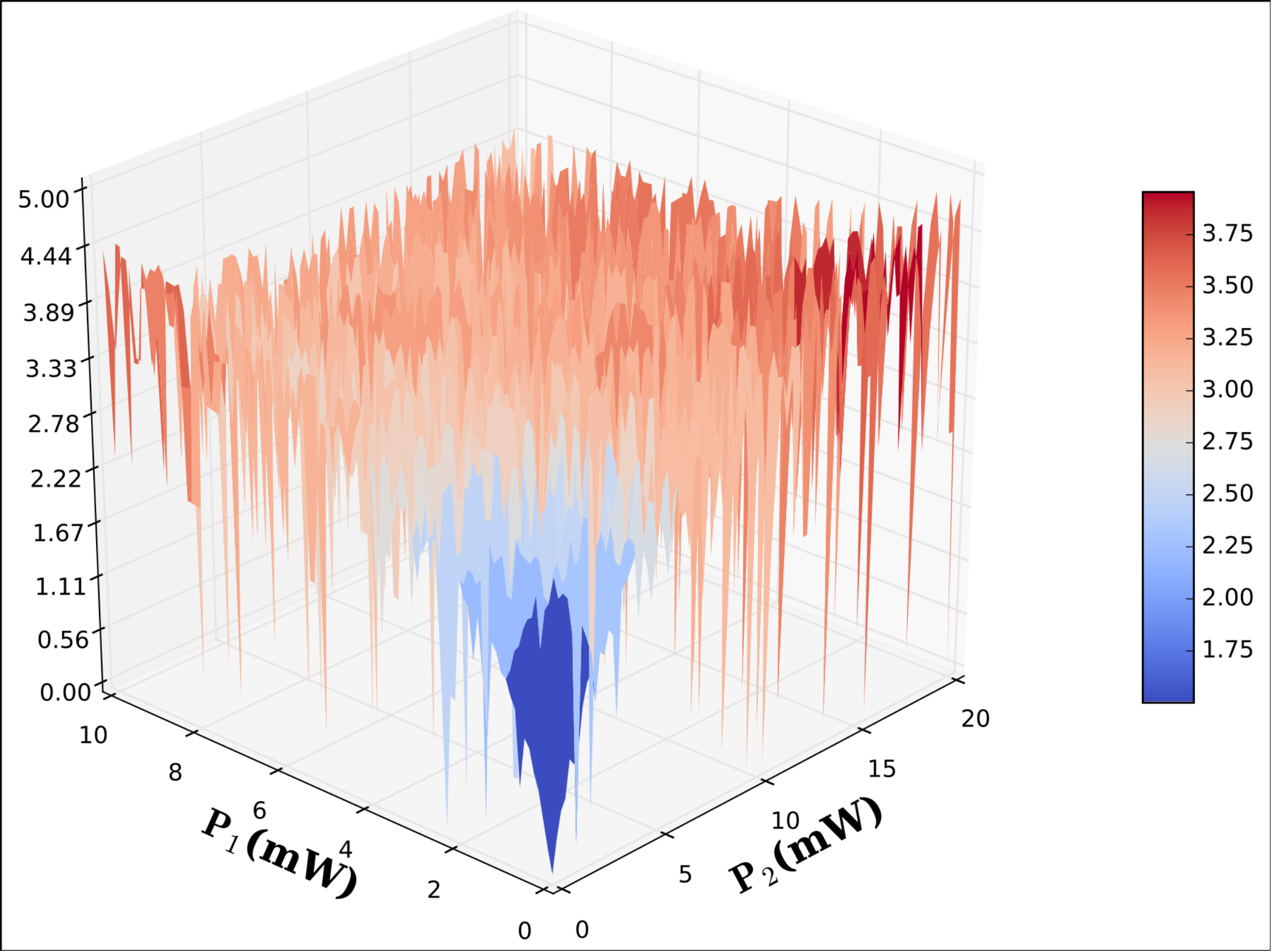}
\caption[width=.3\textwidth]{Global action-value function.}% with maximum at $(9.78,20.0)$.}
\label{fig_result_two_UE}
\end{centering}
\end{figure}

First we will execute our proposed algorithm for $\beta=0.3$. According to the optimal solution, $(0,P_{2,max})$ is the optimal solution. According to Q-CoPA, the SBSs will choose the powers that maximizes the global Q-function. The learned global Q-function, $\mathbf{Q}\left(P_1,P_2\right)$, is plotted in Fig.~\ref{fig_result_two_UE} with maximum value at $P_1=0$ and $P_2=P_{2,max}$, which is optimal. %According to the global function, the maximum transmitters will choose $P_1=9.78$ mW, and $P_2=20.0$ mW as the maximum of the Q-function, which is almost optimal.
\vspace{-5pt}

\begin{figure}[h]
\begin{centering}
\includegraphics[width=0.95\columnwidth]{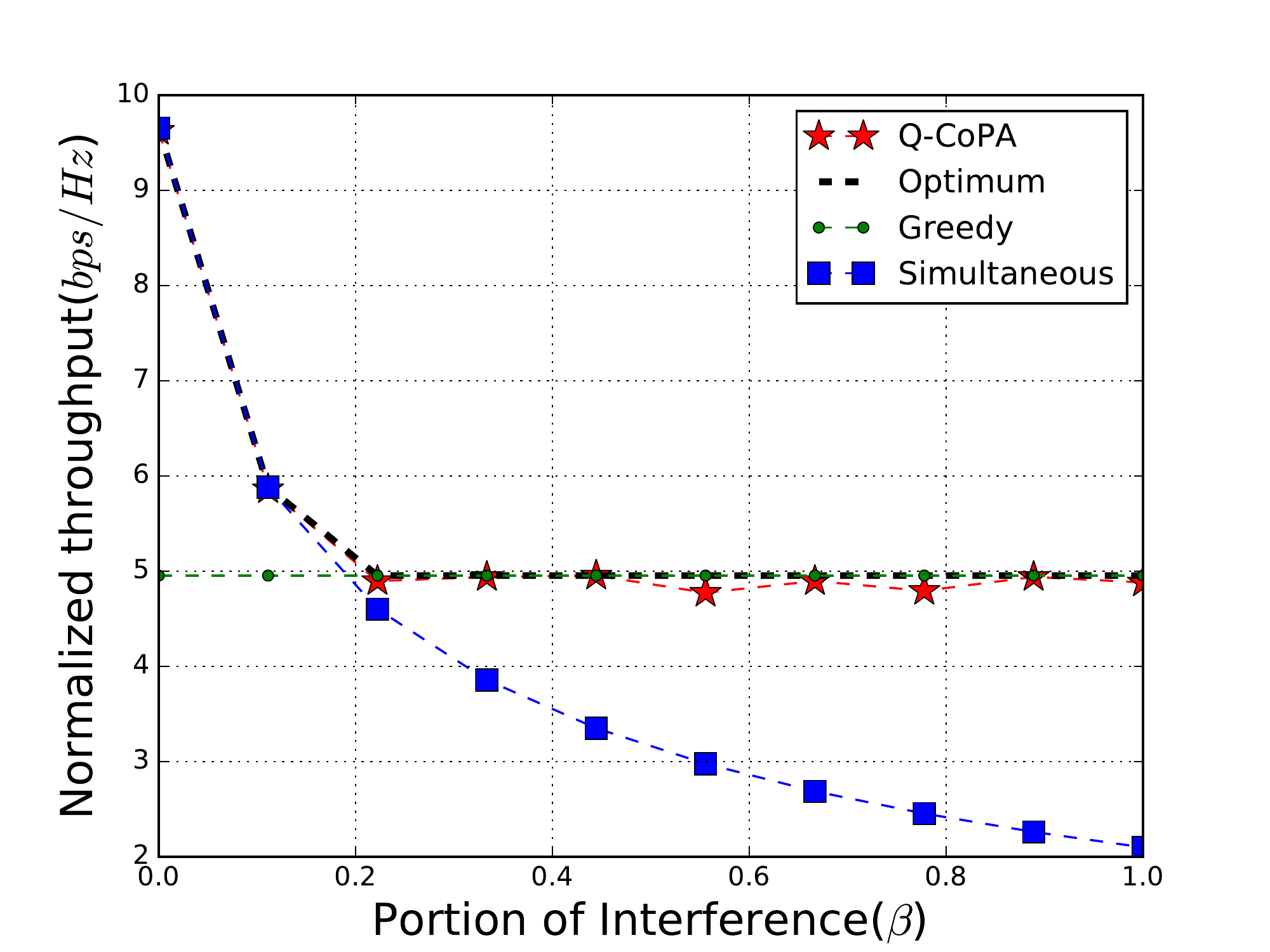}
\caption[width=.5\columnwidth]{\small {Normalized throughput versus portion of interference ($\beta$).}}
\label{fig_total_beta}
\end{centering}
\end{figure}

In Fig.~\ref{fig_total_beta}, the solution of the power allocation for different values of the portion of interference between two channels, $\beta \in \left[0,1\right]$, is plotted. The greedy approach is defined to allocate full power to the transmitter with higher peak power, and zero to the other one. The simultaneous allocation is defined to use maximum power at both transmitters. According to Fig.~\ref{fig_total_beta}, the Q-CoPA finds the optimal solution for all values of $\beta$.

\vspace{-10pt}

%1. I AM CONFUSED WE TALK ABOUT A DENSE NETWORK WE HAVE ONLY TWO. 

%2. WE HAVE NOT DISCUSSED OPTIMALITY UNTIL THE END BUT WE DISCUSS IT IN TERMS OF SIMULATIONS. THAT IS NO SUFFICIENT. 

%\section{Complexity Analysis}
%\vspace{-5pt}
\section{Conclusion} \label{sec_con}
In this paper we used message-passing and variable elimination to coordinate the power allocation in order to maximize a common goal in an interference-limited network. The proposed solution is based on Q-\textit{learning}, and does not need to know the model of the system, hence, it adapts itself if the architecture or number of SBSs in the network changes. Another advantage of this method is that the Q-functions are learned by just measuring the SINR value at each node (radio measurement), while the optimal solution depends on the channel estimation, for example values of $g_1$ and $g_2$ in the simulation in the section~\ref{sec_sim}.

The variable elimination algorithm is exact, so as long as the local Q-functions' action set covers all interfering SBSs, the proposed solution is optimal. Although, when each node of the CG gets densely connected, i.e., the size of action set of local Q-function grows, for the sake of computational complexity we need to approximate local Q-functions' action set with smaller sets, which results in suboptimal solution. Therefore, the proposed solution is suitable for indoor applications, or networks in which the number of interferes is low. As the future work the authors will explore the edge-based decomposition to support outdoor networks and highly dense CGs.

%In the proposed work, the system model works in mmWave, which brings low number of interfering neighbors.The proposed method, is suitable for indoor applications, which there is limited number of BSs, and the architecture of the network can be sporadically changed. In case of LTE, the VE will impose high computation. In that case edge-based decompositions should be integrated, which is part of the future work for us.
\bibliographystyle{IEEEtran}
\bibliography{IEEEabrv,ref_nourl}

\end{document}